\documentclass{lhep}       

\journal{LHEP-519}%Letters in High Energy Physics}
\jyear{2024}
\received{xx January 2018}
\published{xx March 2018}

\def\zzllvvjj{{\ensuremath{ZZ(\rightarrow\ell\ell\nu\nu)jj}}}
\def\zgvvgjj{{\ensuremath{Z\gamma(\rightarrow\nu\nu\gamma)jj}}}
\def\wzlvlljj{{\ensuremath{WZ(\rightarrow\ell\nu\ell\ell)jj}}}
\def\wglvgjj{{\ensuremath{W\gamma(\rightarrow\ell\nu\gamma)jj}}}
\newcommand{\f}[1]{\ensuremath{f_\text{#1}/\Lambda^4}}
\begin{document}

\title{Composite Effective Field Theory Signal from Anomalous Quartic Gauge Couplings for $\zzllvvjj$ and $\zgvvgjj$ Productions}

\author{Artur E. Semushin\auno{1,2}, Evgeny Yu. Soldatov\auno{1}}
\address{$^1$National Research Nuclear University MEPhI, Moscow, Russia}
\address{$^2$A.Alikhanyan National Science Laboratory (Yerevan Physics Institute), Yerevan, Armenia}

\begin{abstract}
Parameterization of heavy effects beyond the Standard Model is available using higher-dimension operators of the effective field theory and their Wilson coefficients, where their values are not known. Experimental sensitivity to the Wilson coefficients can be significantly changed in case of the usage of composite anomalous signal, which contains anomalous contributions from background processes in addition to the conventional ones from the signal process. In this work, this approach is applied to the search for anomalous quartic gauge couplings with seven EFT operators in the electroweak production of $\zzllvvjj$ and $\zgvvgjj$ in $pp$ collisions. For the majority of coefficients, sensitivity in the former channel is smaller than that in the latter one. However, it is shown, that composite anomalous signal affects $\zzllvvjj$ production stronger than $\zgvvgjj$ production, making sensitivities closer. One-dimensional limits on the Wilson coefficients are changed up to 27.3\% and 9.7\% due to the background anomalous contributions in $\zzllvvjj$ and $\zgvvgjj$ productions, respectively.
\end{abstract}

\maketitle


\section{Introduction}
The Standard Model (SM) is a theory of elementary particle physics that describes experimental data in high energy physics well~\cite{Erler:2019hds}. The main aim of such modern experiments is to find out significant deviations from the SM (new physics) in order to determine the correct way for its extension. This extension is expected to explain all or some of the theoretical and experimental issues of the SM, such as gravitational interaction, dark matter and dark energy, and hierarchy problem.

Direct searches for the new particles gave no results at the moment~\cite{Rappoccio:2018qxp, ParticleDataGroup:2022pth}, which indicates a growth of prospects of the indirect approach. It is based on looking for anomalous interactions of the already known particles. An effective field theory (EFT)~\cite{Weinberg:1978kz, Degrande:2012wf} provides a convenient way to parameterize heavy new physics at the current experimental energy scale as anomalous couplings of the SM particles. Its Lagrangian has the following form:
\begin{equation}
    \mathcal{L} = \mathcal{L}_\text{SM} + \sum\limits_{d>4} \mathcal{L}^{(d)}, \quad \mathcal{L}^{(d)} = \sum\limits_i \frac{f_i^{(d)}}{\Lambda^{d-4}} \mathcal{O}_i^{(d)},
\end{equation}
where $\mathcal{L}_\text{SM}$ is dimension-four SM Lagrangian and $\mathcal{L}^{(d)}$ is a dimension-$d$ addition, which is the sum over all dimension-$d$ EFT operators $\mathcal{O}_i^{(d)}$. Each operator is constructed out of the SM fields without gauge symmetry breaking and accompanied by a Wilson coefficient $f_i^{(d)}/\Lambda^{d-4}$, where $\Lambda$ is the scale of new physics. In this work bosonic couplings are studied; therefore, odd-dimension operators are not valid since they contain fermion fields. 

Anomalous quartic gauge couplings (aQGCs) are usually studied using dimension-eight operators, since dimension-six ones do not predict genuine aQGCs~\cite{Eboli:2016kko, Almeida:2020ylr}. Basis of such operators contains three types of $CP$-conserving operators. This work uses a subset of the basis, which includes three T-family operators, constructed from the field strengths,
\begin{align}
    & \mathcal{O}_\text{T0} = \text{Tr}\, \left[ \hat{W}_{\mu\nu} \hat{W}^{\mu\nu} \right] \times \text{Tr}\, \left[ \hat{W}_{\alpha\beta} \hat{W}^{\alpha\beta} \right], \\
    & \mathcal{O}_\text{T1} = \text{Tr}\, \left[ \hat{W}_{\alpha\nu} \hat{W}^{\mu\beta} \right] \times \text{Tr}\, \left[ \hat{W}_{\mu\beta} \hat{W}^{\alpha\nu} \right], \\
    & \mathcal{O}_\text{T5} = \text{Tr}\, \left[ \hat{W}_{\mu\nu} \hat{W}^{\mu\nu} \right] \times B_{\alpha\beta} B^{\alpha\beta},
\end{align}
three M-family operators, constructed from both field strengths and Higgs doublets,
\begin{align}
    & \mathcal{O}_\text{M4} = \left[ (D_\mu \Phi)^\dag \hat{W}_{\beta\nu} D^\mu \Phi \right] \times B^{\beta\nu}, \\
    & \mathcal{O}_\text{M5} = \left[ (D_\mu \Phi)^\dag \hat{W}_{\beta\nu} D^\nu \Phi \right] \times B^{\beta\mu} + \text{h.c.}, \\
    & \mathcal{O}_\text{M7} = (D_\mu \Phi)^\dag \hat{W}_{\beta\nu} \hat{W}^{\beta\mu} D^\nu \Phi,
\end{align}
and one S-family operator, constructed from the Higgs doublet derivatives only,
\begin{align}
    & \mathcal{O}_\text{S0} = \left[ (D_\mu \Phi)^\dag D_\nu \Phi \right] \times \left[ (D^\mu \Phi)^\dag D^\nu \Phi \right].
\end{align}

If some physical process can be produced via an aQGC and the Lagrangian is parameterized by one EFT operator, its squared amplitude can be written as
\begin{equation}
    \begin{aligned}
    |\mathcal{A}|^2 = & \; |\mathcal{A}_\text{SM}|^2 + \left(f/\Lambda^4\right) 2\text{Re} \left(\mathcal{A}_\text{SM}^\dag \mathcal{A}_\text{BSM}\right) \\ & + \left(f/\Lambda^4\right)^2 |\mathcal{A}_\text{BSM}|^2. 
    \end{aligned} \label{eq:decomp}
\end{equation}
It contains three terms: SM, interference (linear) and quadratic. Non-SM terms are referred to as the anomalous contributions.

EFT is used in the analyses of experimental data to set the limits on the Wilson coefficients~\cite{ATLAS:2017vqm, CMS:2020ioi, ATLAS:2020nlt, ATLAS:2022nru}. Often for this purpose, only the signal process is decomposed according to Eq.~\eqref{eq:decomp}, whereas background processes are assumed to have the SM term only. Actually, background processes are also affected by nonzero Wilson coefficients and can have significant anomalous contributions to the analysis signal region. Therefore, usage of the anomalous signal, composed of signal and background anomalous contributions, can lead to the significant changes in the limits on Wilson coefficients. Previously this methodology was tested for $\zgvvgjj$ and $\wglvgjj$ productions and four aQGC operators. Maximum improvement of the limits was found at 9.1\% for $\f{M2}$ coefficient~\cite{Semushin:2022gbm, Semushin:2022amw}. In this work composite anomalous signal is examined for $\zzllvvjj$ and $\zgvvgjj$ productions. The latter one is studied using another event selection compared to the previous studies.

\section{Physical model}
Vector boson scattering (VBS) processes are the most sensitive to the aQGCs. In $pp$ collisions VBS processes are an inherent part of the electroweak (EWK) production of two vector bosons and two jets. However, it is possible to enhance VBS production using special kinematic cuts. In this work, VBS production of $\zzllvvjj$ and $\zgvvgjj$ processes is studied. These processes are rare, the first one has not been observed in the considered decay channel yet~\cite{ATLAS:2020nlt}, whereas the second one has been observed using data from LHC Run~II~\cite{ATLAS:2021pdg}.

\begin{figure*}[h]
\centering
\includegraphics[width=0.6\linewidth]{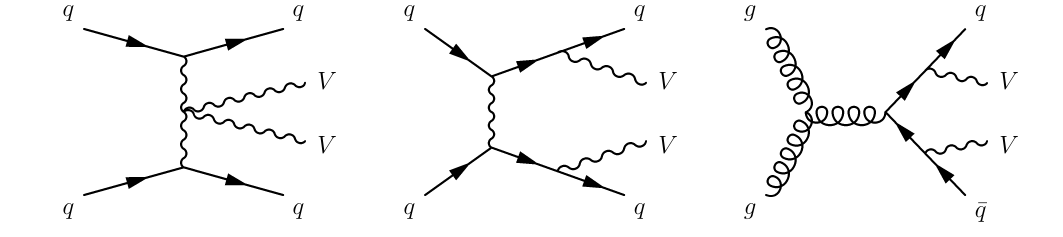}
\caption{Example Feynman diagrams for VBS (left), EWK non-VBS (center), and QCD (right) productions of two vector bosons with two associated jets.}
\label{fig:diagrams}
\end{figure*}

The main background to the EWK production of $\zzllvvjj$ and $\zgvvgjj$ is partially QCD production of the same final states. Example Feynman diagrams for VBS, EWK non-VBS and QCD productions of two vector bosons and two jets are presented in Figure~\ref{fig:diagrams}. Large backgrounds for EWK $\zzllvvjj$ and $\zgvvgjj$ productions come from $\wzlvlljj$ and $\wglvgjj$ production respectively, where one lepton is produced outside of the detector acceptance. In $\zgvvgjj$ channel production of $tt\gamma$ is accounted as an additional background. The sum of all other backgrounds in both channels is not larger than the SM (EWK and QCD) contribution from the signal process, which is used as its estimation. This includes nonresonant background to $\zzllvvjj$ production, minor backgrounds and backgrounds usually estimated directly from data, e.g. the ones caused by the misidentification of the electron as a photon~\cite{Kurova:2023txz}.
\begin{figure*}[h!]
\centering
\includegraphics[width=0.3\linewidth]{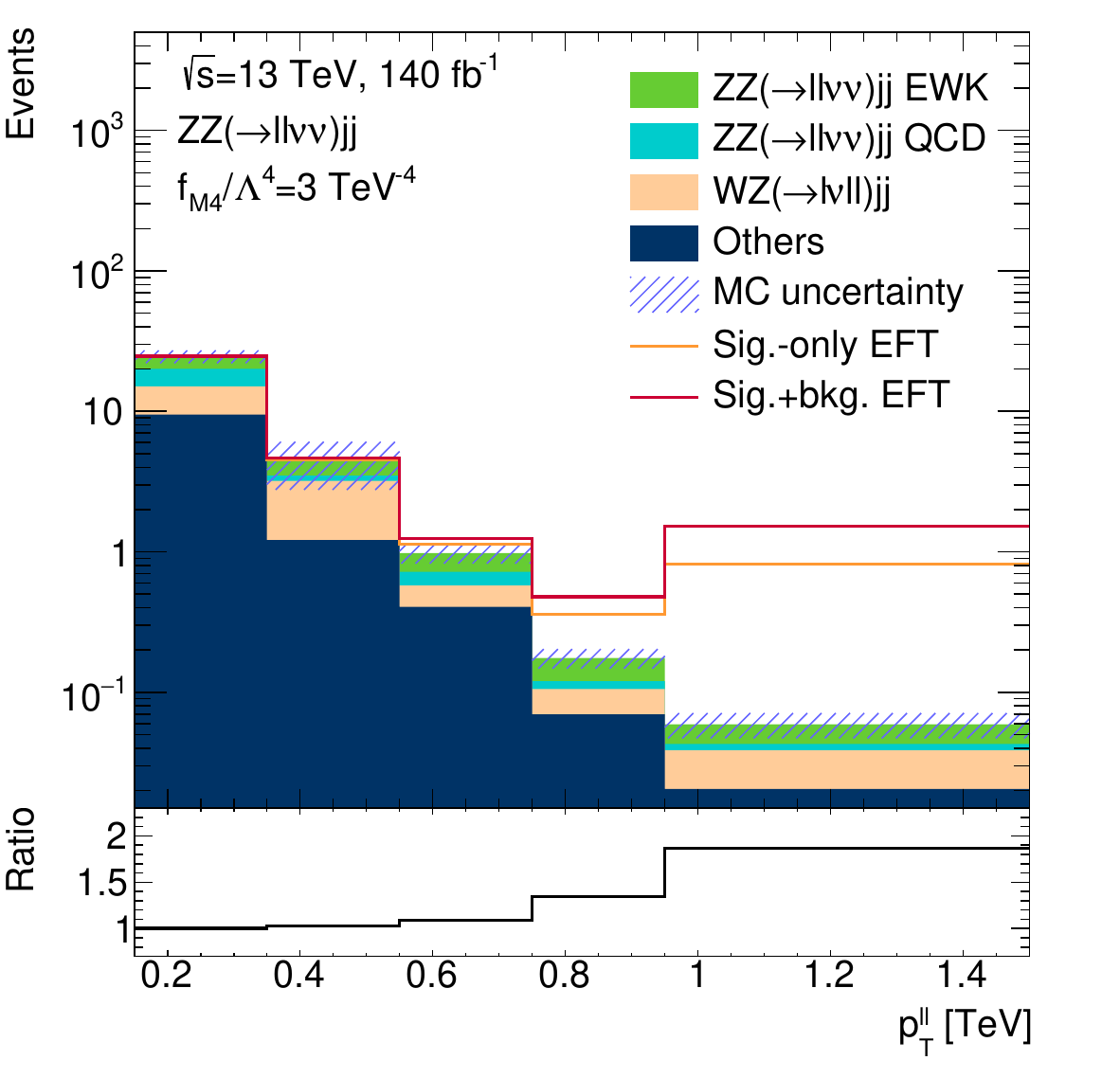}
\hspace{2cm}
\includegraphics[width=0.3\linewidth]{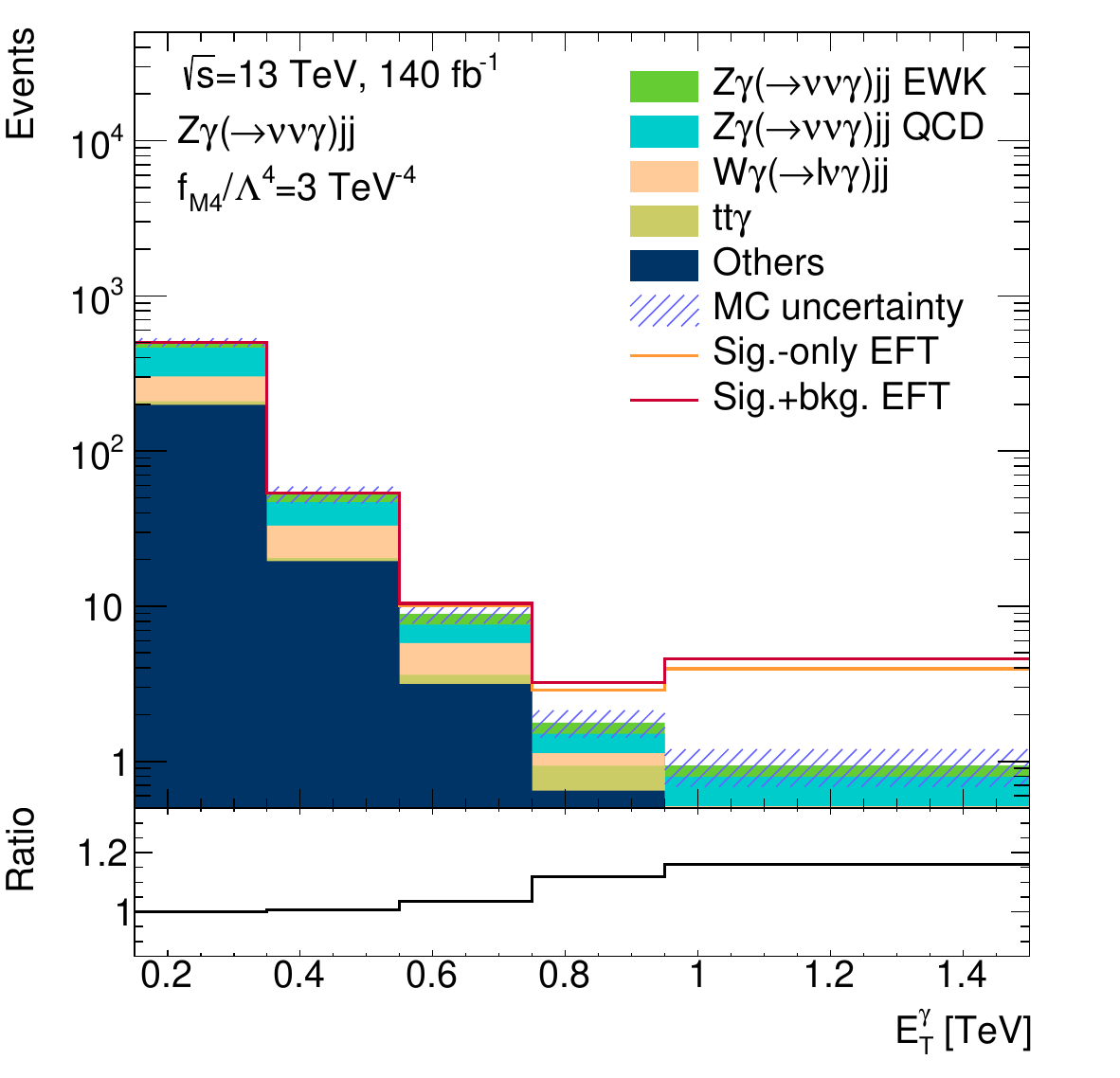}
\caption{Distribution by $p_\text{T}^{\ell\ell}$ for $\zzllvvjj$ channel (left) and distribution by $E_\text{T}^\gamma$ for $\zgvvgjj$ channel (right). Orange (red)~line represents EFT prediction for nonzero $\f{M4}$ coefficient in case BSM contribution is accounted from signal only (signal and background). Ratio of red and orange lines is presented in the lower panel.}
\label{fig:distributions}
\end{figure*}

In order to model the physical processes, the Monte Carlo event generator \texttt{MadGraph5\_aMC@NLO}~\cite{Alwall:2014hca} is used in this work for parton-level calculations. Hadronization, parton shower and underlying event are simulated using \texttt{Pythia8}~\cite{Sjostrand:2014zea}. The response of the typical particle detector operating at the LHC for the simulated events is parameterized using \texttt{Delphes3}. Calculations of the event yields are made using integrated luminosity of 140~fb$^{-1}$, collected by the ATLAS detector during LHC Run~II~\cite{ATLAS:2022hro}.

After the detector simulation, VBS-enhanced event selection criteria based on the ATLAS studies of $\zzllvvjj$ and $\zgvvgjj$ productions~\cite{ATLAS:2020nlt, ATLAS:2022nru} are applied, summarized in Table~\ref{tab:selection}\footnote{The same coordination system as in the ATLAS studies~\cite{ATLAS:2020nlt, ATLAS:2022nru} is used in this work. $y$ is the rapidity, and $\gamma$-centrality$=|(y_\gamma-(y_{j_1}+y_{j_2})/2)/(y_{j_1}-y_{j_2})|$. Transverse momentum is measured in the plane that is transverse with respect~to the beam pipe. $E_\text{T}^\text{miss}$ is the magnitude of the missing transverse momentum $p_\text{T}^{\text{~miss}}$, and event-based $E_\text{T}^\text{miss}$-significance $=E_\text{T}^\text{miss}/\sqrt{\sum E_\text{T}}$.}. Both ATLAS studies use object-based missing transverse energy significance ($E_\text{T}^\text{miss}$-significance). In this work, it was changed to the event-based one~\cite{ATLAS:2018uid} for the simplicity, and the cutting value was chosen to obtain similar SM signal efficiency as in the ATLAS studies. Moreover, for the $\zzllvvjj$ channel, a cut $p_\text{T}^{\ell\ell}>150$ GeV was added since the sensitivity to aQGCs is larger at high $p_\text{T}^{\ell\ell}$.
\begin{table}[t ]
\tbl{Event selection criteria for $\zzllvvjj$ and $\zgvvgjj$ channels.\label{tab:selection}}{%
\begin{tabular}{p{0.45\linewidth}p{0.45\linewidth}}
\hline
$\boldsymbol{\zzllvvjj}$ & $\boldsymbol{\zgvvgjj}$ \\
\hline
2 leptons of the same flavors and opposite charges &1 photon,\; $E_\text{T}^\gamma>150$ GeV \\
$80<m_{\ell\ell}<100$ GeV & $E_\text{T}^\text{miss}>120$ GeV \\
$E_\text{T}^\text{miss}\text{-signif.}>8$ GeV$^{1/2}$ & $E_\text{T}^\text{miss}$-signif.$>10.5$ GeV$^{1/2}$ \\
$p_\text{T}^{\ell_1}>30$ GeV,\; $p_\text{T}^{\ell_2}>20$ GeV & Lepton veto \\
$N_\text{jets} \geq 2$,\; $m_{jj}>400$ GeV & $N_\text{jets} \geq 2$,\;
$m_{jj}>300$ GeV \\
$p_\text{T}^{j_1}>60$ GeV,\; $p_\text{T}^{j_2}>40$ GeV & $p_\text{T}^j>50$ GeV \\
$|y_{j_1}-y_{j_2}|>2$ & $\gamma\text{-centrality}<0.6$ \\
$N_{b\text{-jet}}=0$ & $|\Delta \varphi (\gamma, \vec{p}_\text{T}^{\text{~miss}})|>0.4$ \\
$p_\text{T}^{\ell\ell}>150$ GeV & $|\Delta \varphi (j_{1,2}, \vec{p}_\text{T}^{\text{~miss}})|>0.3$ \\
\hline
\end{tabular}}
\end{table}

In addition to the signal process, some backgrounds can also have BSM contributions to the signal region. In this work, such backgrounds are $\wzlvlljj$ and $\wglvgjj$ for $\zzllvvjj$ and $\zgvvgjj$ productions respectively. Additionally, BSM contributions from backgrounds $WW(\rightarrow\ell\nu\ell\nu)jj$, $ZZ(\rightarrow 4\ell)jj$ and $Z\gamma(\rightarrow\ell\ell\gamma)jj$ were considered and found to be negligible; therefore they are not used in this work. Figure~\ref{fig:distributions} presents distributions for the variables sensitive to aQGCs, $p_\text{T}^{\ell\ell}$ for $\zzllvvjj$ production and $E_\text{T}^\gamma$ for $\zgvvgjj$ production. SM contributions are shown using filled histograms, whereas solid lines represent EFT predictions for one Wilson coefficient. Usage of composite anomalous signal, i.e. signal and background anomalous contributions, increases predicted event yields, which lead to the improvement of the limits on the Wilson coefficient.

A possible issue in using the background BSM contributions can come from the fact that usually significant backgrounds are estimated using data, e.g., from the special control regions, enriched by the events of a specific background type. In this case, possible background BSM contributions can be included in its estimation. However, this can be avoided, if the control region is not sensitive to the aQGCs. Often this condition is satisfied since the control regions are usually enriched by the background SM events~\cite{ATLAS:2022nru}. Therefore, usually composite anomalous signal method can be used in real analyses.

\section{Results}
In this work, limits on the Wilson coefficients are set with the frequentist statistical method. Test statistic based on the likelihood ratio is used, and its distribution is assumed to be asymptotic; i.e., it corresponds to the chi-squared distribution with one or two degrees of freedom for one-dimensional limits or two-dimensional contours~\cite{Wilks:1938dza, Cowan:2010js}. The likelihood function is constructed as a product of Poisson distributions for each bin from Figure~\ref{fig:distributions} and Gaussian constraints for the nuisance parameters. In addition to the Monte Carlo modeling uncertainties, an additional systematic uncertainty of 20\% is applied, which is typical for such kind of analyses~\cite{ATLAS:2022nru}. However, this does not make a significant contribution to the results.

Table~\ref{tab:1Dresults} summarizes one-dimensional limits on the Wilson coefficients set with and without taking into account BSM contributions from backgrounds $\wzlvlljj$ and $\wglvgjj$ for $\zzllvvjj$ and $\zgvvgjj$ productions respectively. It should be emphasized, that the limits on $\f{S0}$ are set only using the $\zzllvvjj$ channel since the corresponding operator does not affect quartic couplings with photons. The limits are set for two separate channels as well as for their combination. Improvement of the confidence interval is also presented and is significant even in the case of combination. Improvement of the limits on $\f{T0}$ and $\f{T5}$ in $\zgvvgjj$ channel is smaller than the one obtained in~\cite{Semushin:2022amw}, which shows the dependence of the method on the event selection.
\begin{table}[h!]
\tbl{One-dimensional limits on the Wilson coefficients set with and without composite anomalous signal. Improvement of the confidence interval obtained with this method is shown in the last column.\label{tab:1Dresults}}{%
\begin{tabular}{cccc}
\hline
Coef. & Sig.-only EFT & Sig.+bkg. EFT & Impr. \\
\hline
\multicolumn{4}{c}{$\zzllvvjj$} \\
\hline
$\f{T0}$ & [-0.238; 0.224] & [-0.229; 0.216] & 3.7\% \\
$\f{T1}$ & [-0.307; 0.305] & [-0.259; 0.263] & 14.8\% \\
$\f{T5}$ & [-0.586; 0.562] & [-0.557; 0.534] & 4.9\% \\
$\f{M4}$ & [-4.79; 4.78] & [-3.48; 3.48] & 27.3\% \\
$\f{M5}$ & [-6.89; 6.92] & [-5.04; 5.05] & 26.9\% \\
$\f{M7}$ & [-8.47; 8.47] & [7.15; 7.15] & 15.6\% \\
$\f{S0}$ & [-16.7; 16.7] & [-13.4; 13.4] & 19.6\% \\
\hline
\multicolumn{4}{c}{$\zgvvgjj$} \\
\hline
$\f{T0}$ & [-0.103; 0.093] & [-0.101; 0.092] & 1.0\% \\
$\f{T1}$ & [-0.129; 0.127] & [-0.120; 0.121] & 6.2\% \\
$\f{T5}$ & [-0.097; 0.108] & [-0.095; 0.106] & 2.1\% \\
$\f{M4}$ & [-3.01; 3.03] & [-2.72; 2.73] & 9.7\% \\
$\f{M5}$ & [-2.35; 2.30] & [-2.16; 2.12] & 8.0\% \\
$\f{M7}$ & [-11.2; 11.2] & [-10.3; 10.3] & 8.4\% \\
\hline
\multicolumn{4}{c}{Combination} \\
\hline
$\f{T0}$ & [-0.098; 0.089] & [-0.097; 0.088] & 1.3\% \\
$\f{T1}$ & [-0.123; 0.122] & [-0.113; 0.115] & 7.2\% \\
$\f{T5}$ & [-0.097; 0.107] & [-0.095; 0.105] & 2.1\% \\
$\f{M4}$ & [-2.71; 2.72] & [-2.30; 2.31] & 15.1\% \\
$\f{M5}$ & [-2.29; 2.25] & [-2.07; 2.03] & 9.7\% \\
$\f{M7}$ & [-7.27; 7.28] & [-6.31; 6.31] & 13.3\% \\ 
\hline
\end{tabular}}
\end{table}

\begin{figure*}[h!]
\centering
\includegraphics[width=0.3\linewidth]{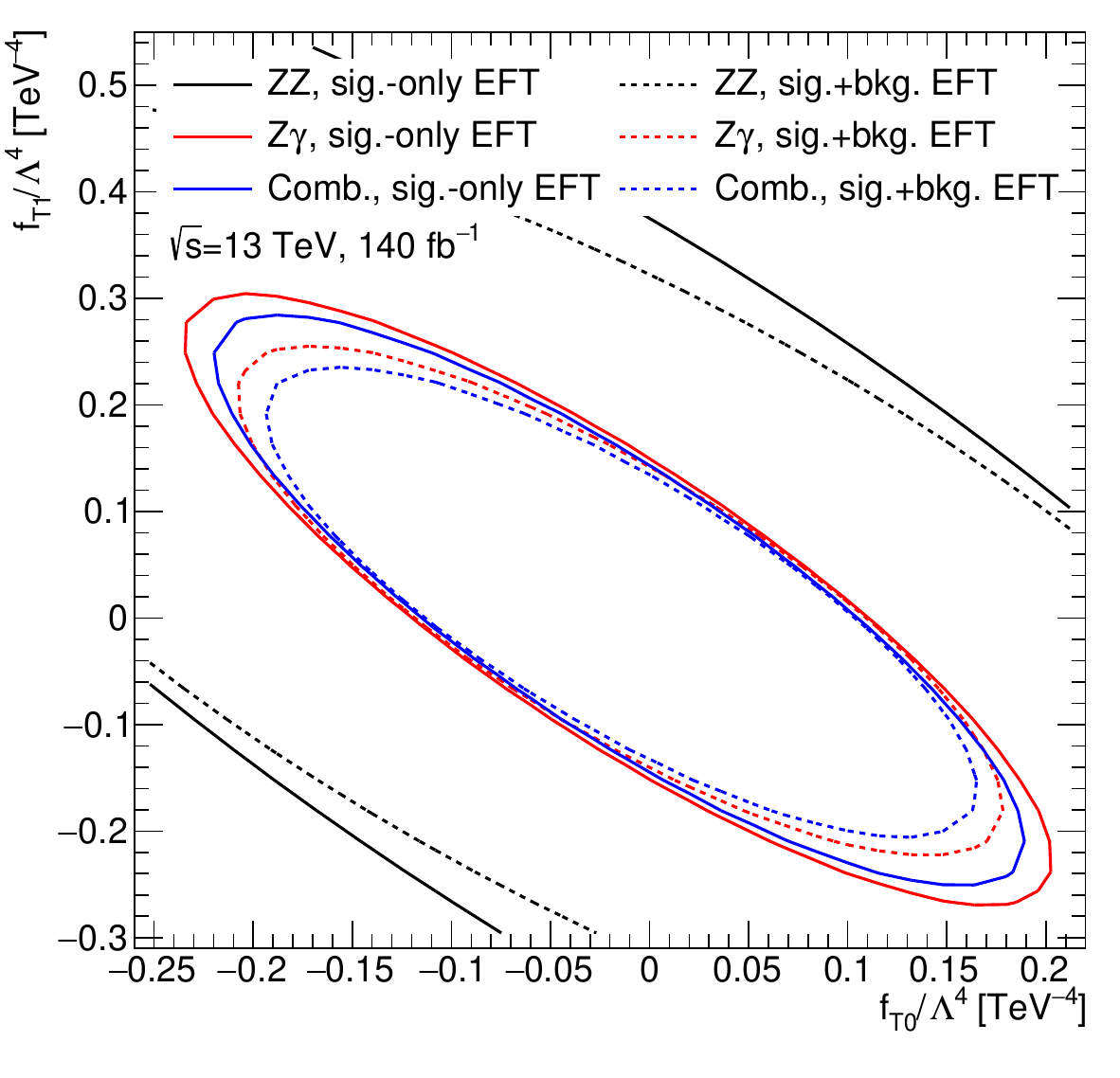}
\hspace{2cm}
\includegraphics[width=0.3\linewidth]{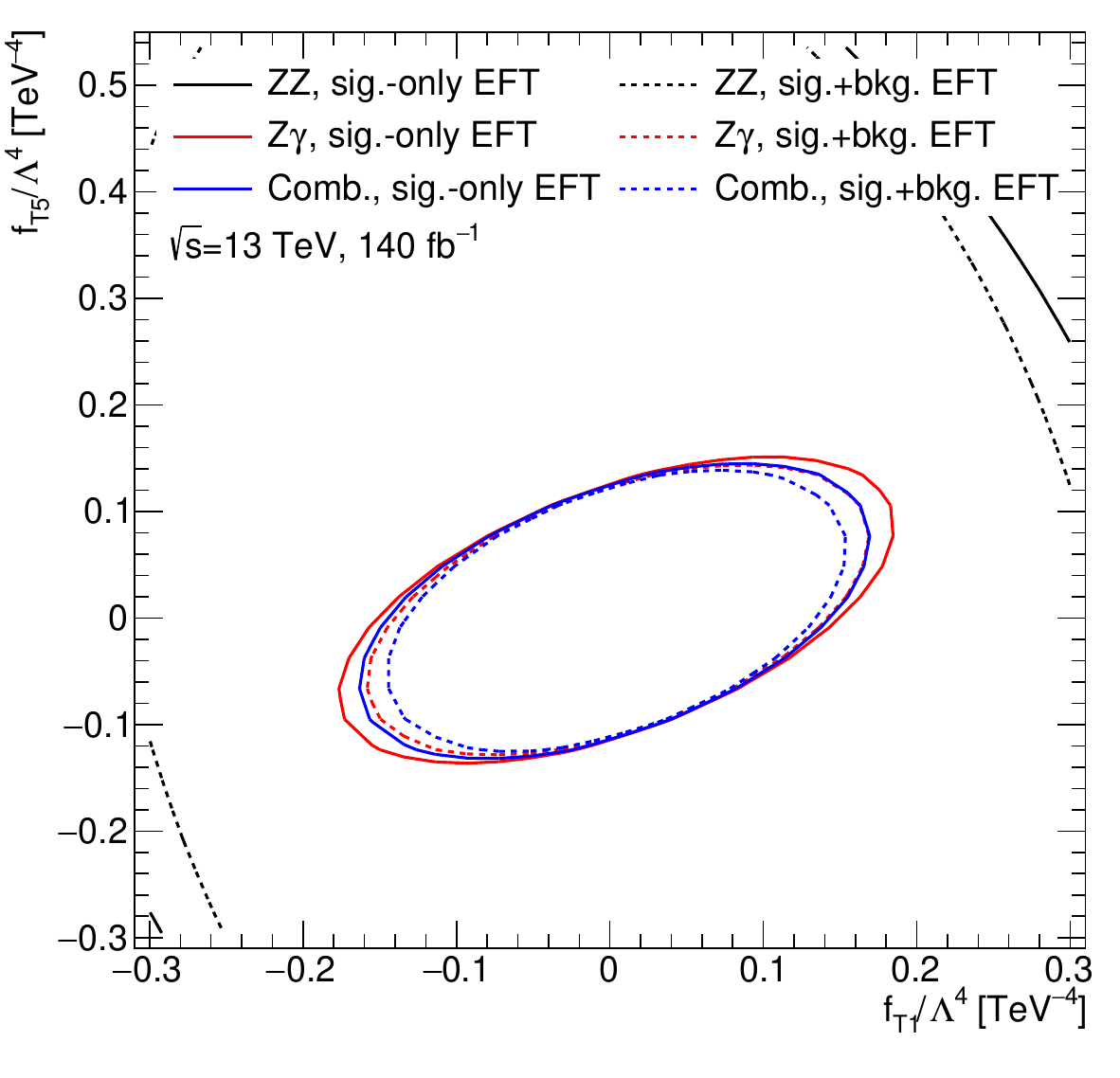}

\includegraphics[width=0.3\linewidth]{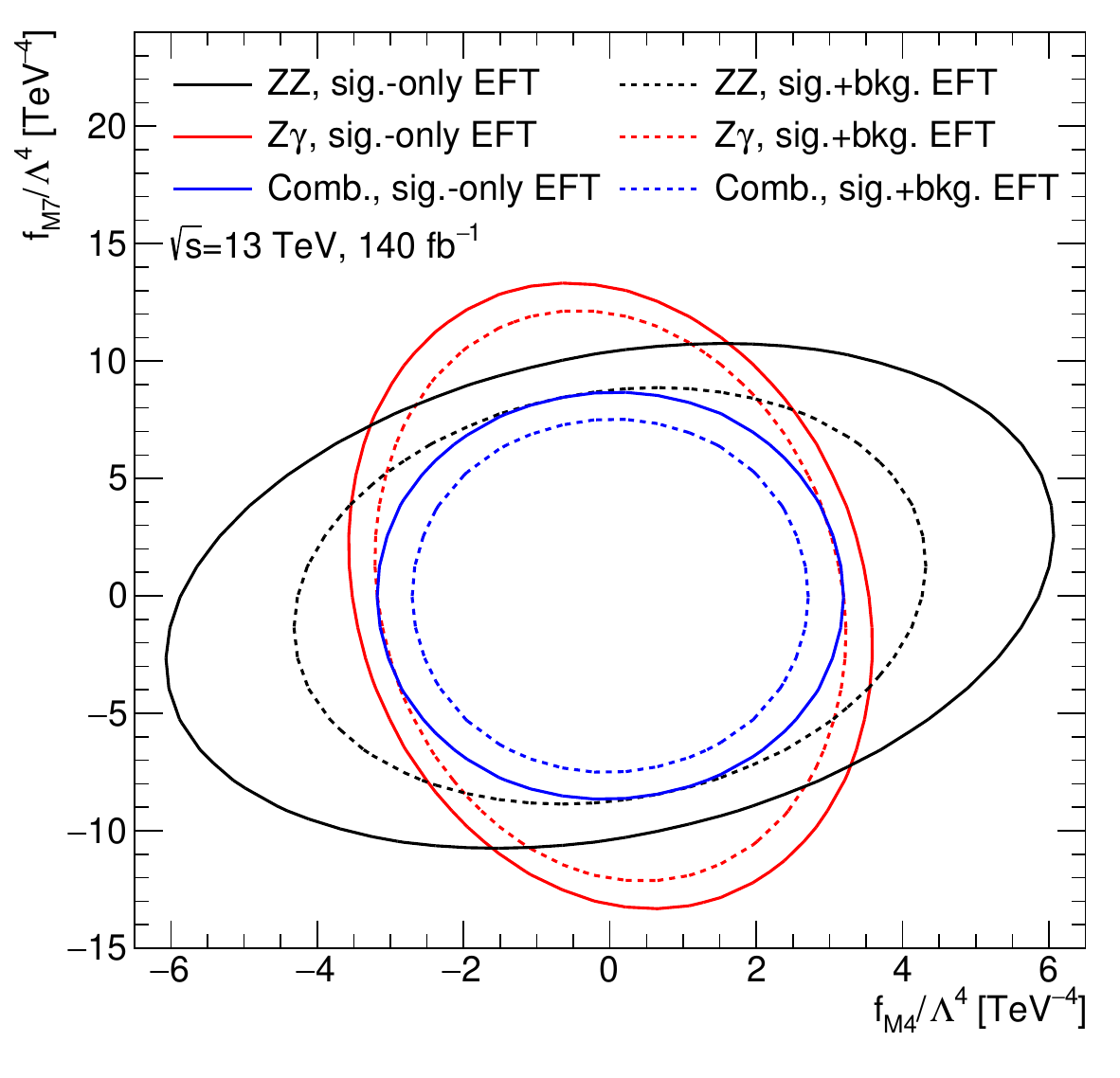}
\hspace{2cm}
\includegraphics[width=0.3\linewidth]{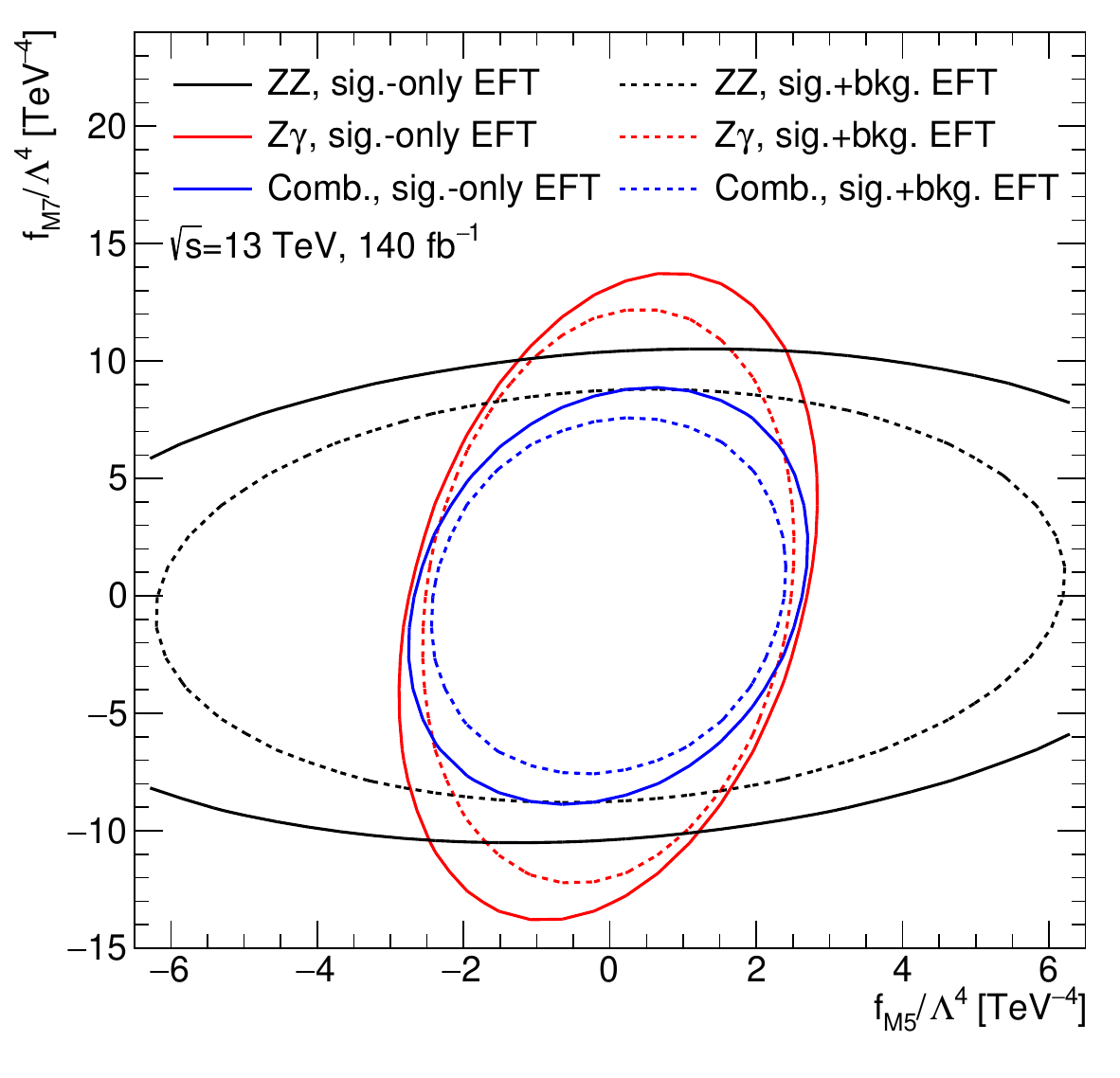}
\caption{Two-dimensional limits on four pairs of the Wilson coefficients. Solid (dashed) lines were obtained without (with) composite anomalous signal. Black, red and blue lines were obtained from the $\zzllvvjj$ channel, $\zgvvgjj$ channel, and their~combination.}
\label{fig:2Dresults}
\end{figure*}

Two-dimensional limits are an additional illustration for the method of composite anomalous signals. Two-dimensional parameterization of the Lagrangian leads to the additional term compared to the decomposition from Eq.~\eqref{eq:decomp}. This term, the so-called cross-term, represents interference between two EFT operators and can significantly change the sensitivity, so studies of two-dimensional limits are also important. In this work, two-dimensional limits are set for four pairs of the Wilson coefficients and presented in Figure~\ref{fig:2Dresults}. The largest improvement of the confidence regions is 40.0\%, 18.6\%, and 26.3\% for $\zzllvvjj$, $\zgvvgjj$, and combined channel, respectively.

\section{Conclusion}
Application of the method of composite anomalous signal to the search for anomalous quartic gauge couplings in $\zzllvvjj$ and $\zgvvgjj$ productions in $pp$ collisions is investigated in this work. In each channel, only one background has significant BSM contributions, and this allows improving one-dimensional limits on the Wilson coefficients up to 27.3\%, 9.7\% and 15.1\% for $\zzllvvjj$ production, $\zgvvgjj$ production, and their combination, respectively. Significant improvement of these limits can lead to a sizeable improvement of the limits on new physics model parameters since model-independent and model-dependent approaches can be connected~\cite{Remmen:2019cyz}. Despite the fact that this work is made for an integral luminosity of 140 fb$^{-1}$, this method is prospective to be used in the future, e.g., in LHC Run III data analyses, since it works independently on the luminosity growth.

\section*{Conflicts of interest}
The authors declare that there are no conflicts of interest regarding the publication of this paper.

\section*{Acknowledgments}
This work was supported by the Russian Science Foundation under grant 21-72-10113.

\bibliographystyle{unsrt}

\end{document}